\documentclass[a4paper, oneside, twocolumn, notitlepage, 10pt]{extarticle_ecoc}
\usepackage{ecoc}

\usepackage{comment}
\usepackage{color}
\usepackage{subfigure}
\usepackage{enumerate}
\usepackage{enumitem}
\usepackage{stfloats}
\usepackage{psfrag}
\usepackage{setspace}
\usepackage{balance}
\usepackage{enumitem}
\usepackage{tikz}
\usepackage{pgfplots,amsfonts}
\usepackage{pgfplotstable}
\usepackage{amsmath,amssymb}
\usepackage{xcolor}
\usepackage{graphicx}
\usepackage{tikz}
\usepackage{pgfplots}
\usepackage{float}
\usepackage{balance}
\usepackage{setspace}
\usepackage{xspace}
\usepackage[normalem]{ulem}

\usepackage{dsfont,empheq}
\usepackage{multicol,multirow}
\usepackage{gensymb}

\usepackage{booktabs, cellspace, hhline}

\usetikzlibrary{shapes}
\usetikzlibrary{spy}
\usetikzlibrary{fit}
\usetikzlibrary{shapes.multipart}
\usetikzlibrary{positioning}

\usetikzlibrary{plotmarks,matrix,chains,scopes,fit,calc,shapes,positioning,decorations,intersections,arrows,backgrounds,shadows}

\newlength\FigureHeight
\newlength\FigureWidth
\definecolor{myDarkGreen}{rgb}{0.00000,0.58824,0.00000}%
\definecolor{uniform}{rgb}{0.00000,0.58824,0.00000}%
\definecolor{AWGNreference}{rgb}{0.00000,0.58824,0.00000}%

\newcounter{lemma}

\newtheorem{exampleplain}{Example}

\newtheorem{definitionplain}{Definition}

\newcommand{\bC}{\boldsymbol{C}}
\newcommand{\bb}{\boldsymbol{b}}

\newcommand{\bs}{\boldsymbol{s}}

\pgfplotsset{compat=1.14}
\usepackage{wrapfig}
\usepackage[acronym,nomain]{glossaries}

\newacronym{CUT}{CUT}{channel under test}

\newacronym{KK}{KK}{Kramers-Kronig}
\newacronym{CSPR}{CSPR}{carrier-to-signal power ratio}
\newacronym{KKRX}{KKRx}{Kramers-Kronig Receiver}
\newacronym{SSBI}{SSBI}{signal-signal beat interference}

\newacronym{GS}{GS}{geometric shaping}
\newacronym{PS}{PS}{probabilistic shaping}
\newacronym{DSP}{DSP}{digital signal processing}
\newacronym{MIMO}{MIMO}{multiple-input multiple-output}
\newacronym{TDE}{TDE}{time domain equalizer}
\newacronym{FDE}{FDE}{frequency domain equalizer}
\newacronym{LMS}{LMS}{least means square}
\newacronym{DDLMS}{DD-LMS}{decision directed least means square}
\newacronym{FFE}{FFE}{feed-forward equalizer}
\newacronym{FBE}{FBE}{feedback equalizer}
\newacronym{BPS}{BPS}{blind phase search}

\newacronym{SMF}{SMF}{single-mode fiber}
\newacronym[plural=SSMFs]{SSMF}{SSMF}{standard single-mode fiber}
\newacronym[plural=FMFs]{FMF}{FMF}{few-mode fiber}
\newacronym{FMF12}{FMF12}{12 mode FMF}
\newacronym{MMF}{MMF}{multi-mode fiber}
\newacronym{SI}{SI}{step index}
\newacronym{GI}{GI}{graded index}
\newacronym{DCF}{DCF}{dispersion compensated fiber}

\newacronym{SDM}{SDM}{space division multiplexing}
\newacronym{MDM}{MDM}{mode division multiplexed}
\newacronym{WDM}{WDM}{wavelength division multiplexing}
\newacronym{DWDM}{DWDM}{dense wavelength division multiplexing}
\newacronym{LP}{LP}{linear polarized}
\newacronym[plural=MMUXs,firstplural=mode multiplexers]{MMUX}{MMUX}{mode multiplexer}
\newacronym{PL}{PL}{photonic lantern}
\newacronym{3DWG}{3DWG}{3D-waveguide}
\newacronym{MDL}{MDL}{mode dependent loss}
\newacronym{DGD}{DGD}{differential group delay}
\newacronym{DMGD}{DMGD}{differential mode group delay}
\newacronym{QSM}{QSM}{quasi-single-mode}
\newacronym{GIMMF}{GI-MMF}{graded-index multi-mode fiber}

\newacronym{SSB}{SSB}{single side band}
\newacronym{QPSK}{QPSK}{quadrature phase shift keying}
\newacronym{QAM}{QAM}{quadrature amplitude modulation}
\newacronym{RRC}{RRC}{root-raised-cosine}
\newacronym{4D-64PRS}{4D-64PRS}{
four-dimensional 64-ary polarization-ring-switching}
\newacronym{DP}{DP}{dual-polarization}
\newacronym[\glslongpluralkey=states-of-polarization]{SOP}{SOP}{state-of-polarization}
\newacronym{PM}{PM}{polarization-multiplexed}
\newacronym{MD}{MD}{multi-dimensional}

\newacronym{ECL}{ECL}{external cavity laser}
\newacronym{CW}{CW}{continuous wave}
\newacronym[plural=DFBs]{DFB}{DFB}{distributed feedback laser}

\newacronym[plural=DACs]{DAC}{DAC}{digital-to-analog converter}
\newacronym{ADC}{ADC}{analog-to-digital converter}
\newacronym{PRBS}{PRBS}{pseudo-random bit sequence}
\newacronym{LO}{LO}{local oscillator}
\newacronym{EDFA}{EDFA}{erbium-doped fiber amplifier}
\newacronym{MZM}{MZM}{Mach-Zehnder modulator}
\newacronym{DP-MZM}{DP-MZM}{dual-polarization Mach-Zehnder modulator}
\newacronym{ChUT}{ChUT}{channel under test}
\newacronym{WSS}{WSS}{wavelength selective switch}
\newacronym[plural=VOAs]{VOA}{VOA}{variable optical attenuator}
\newacronym[plural=PDCRXs]{PDCRX}{PDCRX}{polarization diverse coherent receiver}
\newacronym{DSO}{DSO}{digital storage oscilloscope}
\newacronym{ASE}{ASE}{amplified spontaneous emission}
\newacronym{PBS}{PBS}{polarization beam splitter}
\newacronym{PD}{PD}{photodiode}
\newacronym{AOM}{AOM}{acousto-optic modulator}
\newacronym{BPD}{BPD}{balanced photo-diode}
\newacronym{OMFT}{OMFT}{optical multi-format transmitter}
\newacronym{DPIQ}{DP-IQM}{dual-polarization IQ-modulator}
\newacronym{ABC}{ABC}{automatic bias controller}
\newacronym{OTF}{OTF}{optical tunable filter}
\newacronym{LSPS}{LSPS}{loop-synchronous polarization scrambler}
\newacronym{OSA}{OSA}{optical spectrum analyzer}

\newacronym{OSNR}{OSNR}{optical signal to noise ratio}
\newacronym{BER}{BER}{bit error rate}
\newacronym{IL}{IL}{insertion loss}

\newacronym{SDFEC}{SD-FEC}{soft-decision forward error correction}
\newacronym{HDFEC}{HD-FEC}{hard-decision forward error correction}
\newacronym{FEC}{FEC}{forward error correction}
\newacronym{LDPC}{LDPC}{low-density parity-check code}

\newacronym{AIR}{AIR}{achievable information rate}
\newacronym{AR}{AR}{achievable rates}
\newacronym{MI}{MI}{mutual information}
\newacronym{GMI}{GMI}{generalized mutual information}
\newacronym{BICM}{BICM}{bit-interleaved coded modulation}

\newacronym{NLI}{NLI}{non-linear interference}

\newacronym{OVNA}{OVNA}{optical vector network analyzer}
\newacronym{NIR}{NIR}{near infrared}
\newacronym{CD}{CD}{chromatic dispersion}
\newacronym{OTDR}{OTDR}{optical time domain reflectometry}
\newacronym{OFDR}{OFDR}{optical frequency domain reflectometry}
\newacronym{GPU}{GPU}{graphics processing unit}
\newacronym{SVD}{SVD}{singular value decomposition}
\newacronym{WGN}{WGN}{white Gaussian noise}
\newacronym{AWGN}{AWGN}{additive white Gaussian noise}
\newacronym{PDL}{PDL}{polarization dependent loss}
\newacronym{SPS}{sps}{samples-per-symbol}
\newacronym{SE}{SE}{spectral efficiency}

\usepackage{arydshln}

\usepackage[ruled,linesnumbered]{algorithm2e}

\begin{document}
\selectlanguage{english}    


\title{Nonlinear Interference Analysis of Probabilistic Shaping vs. 4D Geometrically Shaped Formats}%


\author{
    Bin Chen\textsuperscript{(1)},
    Chigo Okonkwo\textsuperscript{(2)},
    Alex Alvarado\textsuperscript{(2)}
}

\maketitle                  


\begin{strip}
 \begin{author_descr}
 
   \textsuperscript{(1)}School of Computer Science and Information Engineering, Hefei University of Technology, China
   \textcolor{blue}{\uline{bin.chen@hfut.edu.cn}} 
   
 \textsuperscript{(2)}Department  of  Electrical  Engineering, Eindhoven University of Technology, The   Netherlands 

 \end{author_descr}
\end{strip}

\setstretch{1.093}


\begin{strip}
  \begin{ecoc_abstract}
  	Performance trade-offs between linear shaping and nonlinear tolerance of the recently introduced  4D orthant-symmetric 128-ary modulation format are investigated. Numerical simulations show 9.3\% reach increase with respect to the 7b4D-2A8PSK format and probabilistically-shaped 16QAM with short blocklength.

  \end{ecoc_abstract}
\end{strip}

\section{Introduction}
\vspace{-0.5em}
In optical transmission systems, the performance of a given
modulation format is  determined by its tolerance to both
linear \gls{ASE} noise, and 
\gls{NLI} arising from the Kerr effect.
Designing modulation formats which increase 
achievable information rates (AIRs) in the presence of linear
and nonlinear impairments is crucial to achieve higher  capacity  and  longer reach.

NLI depends on the average power of the transmitted signal. One of the most widely used model for NLI is the so-called Gaussian noise (GN) model. The GN model neglects most of the specific
properties of the transmitted signal, including its
underlying modulation format \cite{Poggiolini_JLT2014}. 
However, in the vast majority of recent demonstrations, it has been shown that NLI also depends on the modulation format \cite{SerenaECOC2014,GaldinoECOC2016}. A model considering modulation format dependency was later developed as an improvement to the GN model,  known as enhanced GN (EGN) model \cite{Carena:14,DarJLT2015}.

Signal shaping has recently been widely investigated in optical fibre communications to improve spectral efficiency (SE), and is currently  implemented in commercial products via  probabilistic shaping (PS) \cite{NokiaPSE-V} and geometric shaping (GS) \cite{FujitsuT600}. 
The performance of PS 
 has been examined in theory, simulations, and experiments \cite{ChoECOC2016,TobiasJLT16,Buchali2016,RennerJLT2017,SillekensOFC2018}.
However, PS suffers from rate loss in practical implementations with finite blocklengths \cite{FehenbergerJLT2020}  and  also experiences increased NLI \cite{TobiasJLT16,FehenbergerOFC2020}.  Multidimensional (MD) geometric shaping based on  constant modulus formats is known to offer NLI tolerance  in the nonlinear channel \cite{Kojima2017JLT,BinChenJLT2019}.
In so doing, the multidimensional modulation format ``shapes out" the partial NLI at the expense of losing only partial degrees of freedom \cite{DarISIT2014}.









In this paper, the recently proposed four-dimensional orthant-symmetric 128-ary (4D-OS128) format \cite{ChenArxiv2020}  are studied that was designed  by maximizing generalized mutual information with  the benefit of significantly reducing the optimization searching space.
We observe a performance trade-offs  between linear shaping and nonlinear tolerance, and thus 4D-OS128  outperforms  the corresponding nonlinearity-tolerant geometrically-shaped constellation 7b-4D2A8PSK and PS-16QAM with finite blocklength  at the same SE.

\vspace{-0.7em}
\section{Orthant-Symmetric MD Geometric Shaping}
To  solve  the  multi-parameter  optimization  challenges of MD geometric shaping and also to reduce the transceiver requirements, we proposed to impose an ``orthant symmetry" constraint to the $N$-dimensional modulation format to be designed\cite{ChenArxiv2020}. 
Orthant-symmetric labeled constellations are be generated from any \textit{first-orthant constellation}, where the constellation points  
are obtained by folding the first-orthant points to the remaining orthants  \cite{ChenArxiv2020}. 

\begin{figure}[!b]
    \vspace{-2em}
	\includegraphics[width=0.23\textwidth]{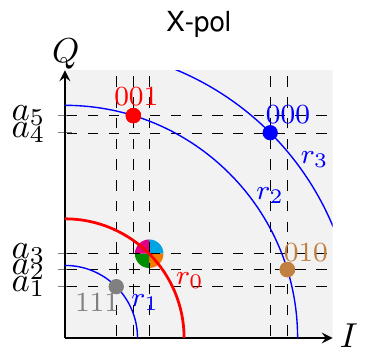}
	\hspace{0em}
	\includegraphics[width=0.23\textwidth]{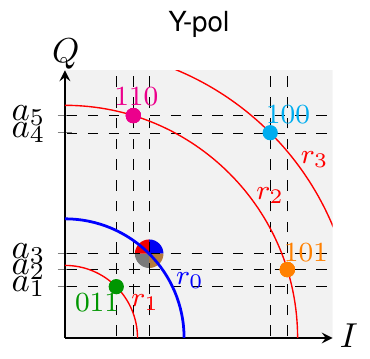}
	\vspace{0em}
	\caption{First orthant of 4D-OS128 modulation formats and associated bit mapping $[b_{j5}, b_{j6}, b_{j7}], j\in\{1,2,\dots,8\}$.}
	\label{fig:firstOrthant}
	\vspace{-1em}
\end{figure}

In this paper, we focus on comparing modulation formats with a \gls{SE} of $m=7$~bit/4D-sym, which consists of $M=2^m$ $N=4$-dimensional points $\bs_i,i\in\{1,2,\dots,128\}$ labeled by 7 bits $\bb_i= [b_1,b_2,\dots,b_7]$.
For the 4D-OS128 format, each orthant contains $2^{m-N}=8$ constellation points. The 8 constellation points in the first orthant 
are denoted by $\boldsymbol{t}_j=\left[t_{j1},t_{j2},t_{j3},t_{j4}\right] \in\mathbb{R}_+^4$ with $j=\{1,2,\dots,8\}$.
The first orthant is labeled by four  binary bits $[b_{j1},b_{j2},b_{j3},b_{j4}]$, which determine one of sixteen orthants.
The remaining three bits $[b_{j5}, b_{j6}, b_{j7}]$ determine one of eight point $\boldsymbol{t}_j$ in the corresponding orthant.
The 2D projections of the first orthant of the 4D-OS128 \cite{ChenArxiv2020} format is shown in Fig. \ref{fig:firstOrthant}, where the  symbols  with  the  same  color belong to the same 4D symbol. Due to the structure of the 4D-OS128 format, the 4D symbols have three energy levels $r_0^2+r_k^2, k\in\{1,2,3\}$ (highlighted as three red or blue circle combinations   in Fig. \ref{fig:firstOrthant}). The  five amplitude values of the the 4D-OS128 format  in Fig. \ref{fig:firstOrthant} optimized for SNR of $9.5$~dB  are $(a_1,a_2,a_3,a_4,a_5)=(0.2875, 0.3834,0.4730,1.1501,1.2460)$.

\begin{figure*}[!tb]
	\includegraphics[width=1\textwidth]{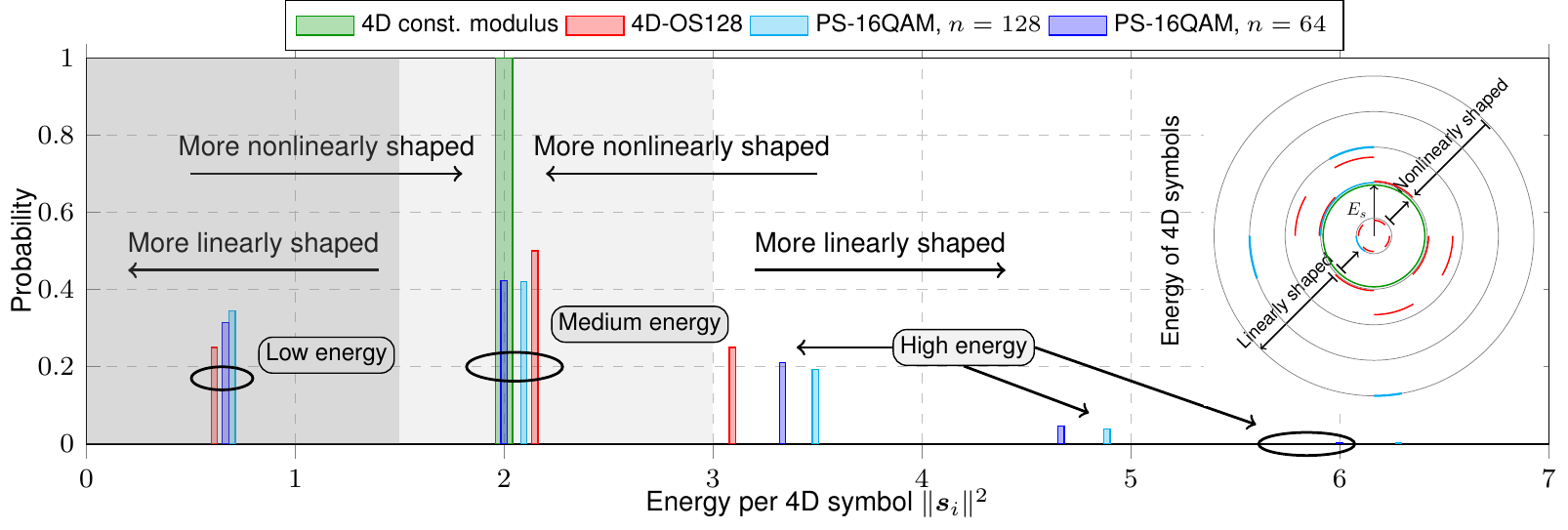}
	\vspace{-1em}
	\caption{Probability of energy per 4D symbol for four \gls{SE} = 7 bits/4D-sym modulations.  The variation of the transmitted symbols'  energy  contributes to the nonlinear interference  noise (NLIN). Inset: 2D geometric representation of energy per 4D symbol.}
		\label{fig:4DEnergy_hist}
\end{figure*}

\vspace{-0.5em}
\section{Probabilistic Amplitude Shaping}
	

In addition to the orthant-symmetric MD modulation format, we also  consider probabilistic amplitude shaping
(PAS) with  quadrature
amplitude modulation (QAM) to achieve the same \gls{SE} of 7~bit/4D-sym.
In this paper, PS-16QAM is generated via   probabilistically shaping   polarization-multiplexed (PM)-16QAM with constant composition distribution matching (CCDM).
To take  finite-length CCDM rate loss into account, the AIRs of PAS for bit-metric decoding (BMD) and for a finite-length CCDM of length $n$ is computed as \cite{2018Tobias_PBDM},
\vspace{-1em}
\begin{equation}
\vspace{-1em}
\text{AIR}_n=[H(\bC)-\sum^{m}_iH(C_i|Y)]-R_{ \text{loss},n},	   
\vspace{-0em}
\end{equation}
where  $\bC=(C_1,\dots,C_m)$ represents the $m$ coded bit levels of the considered modulation format, $H(\cdot)$ denotes entropy, $R_{ \text{loss},n}$ indicate the  finite-length rate loss \cite[eq. (4)]{Bcherer2017arxiv} and $Y$ is the channel output. 

\vspace{-0.5em}
\section{Modulation-dependent NLI}
NLI in the EGN model is effectivel considered as additive white Gaussian noise. 
It is well understood that low energy variations in the signal reduce NLI \cite{GhazisaeidiJLT2017}.
Therefore, choosing symbols with less energy variations and  close to the  envelope of a four-dimensional ball can lead to less NLI, which can be considered as \textit{nonlinear shaping}.
This is in contradiction with Gaussian shaped constellations choosing symbols within the multidimensional balls, 
which is referred as \textit{linear shaping}.


Fig. \ref{fig:4DEnergy_hist} shows the example of \textit{linear shaping} and \textit{nonlinear shaping} by  comparing the  probability distribution of 4D symbol's energy  for three different formats (all normalized to unit energy per polarization): 4D constant-modulus (CM) constellations, 4D-OS128 and PS-16QAM with block length of $n=64$ and $n=128$. 
We can observe that  nonlinear shaping is in contradiction with linear shaping, which  moves the constellation symbols away from the average energy $E_s$ (also see 2D geometric representation of energy per 4D symbol as inset in Fig. \ref{fig:4DEnergy_hist}). 
In addition, the 4D constellation symbols' energy  spread out from the average normalized energy $E_s=2$, which are divided as three groups: low energy symbols, medium energy symbols and high energy symbols. We will show in following section that this 4D energy distribution can induce NLI fluctuation for the nonlinear fibre channel.

\def\arraystretch{0.95}
\begin{table}[!tb]
\caption{Simulation parameters.}
\vspace{-1em}
\begin{center}
\label{tab:syspar}
{\footnotesize
\begin{tabular}{lr}


\toprule
{\bf Parameter name} & {\bf Value}  \\
\midrule
Symbol rate & {41.79~GBd}   \\
Root-raised-cosine roll-off factor & {1}{\%} \\
Channel frequency spacing & {50~GHz}      \\
Center wavelength  & {1550~nm}  \\
Attenuation & {0.21~dB/km} \\
Dispersion parameter & {16.9~ps nm$^{-1}$km $^{-1}$} \\
Nonlinearity parameter & {1.31~W $^{-1}$ km $^{-1}$} \\
\toprule
\end{tabular}
}
\end{center}
\vspace{-2.5em}
\end{table}
\vspace{-0.7em}
\section{Numerical Simulations}
Split-step Fourier method simulations with a step size of 0.1~km were performed to compare the modulation formats and predict system performance. The simulation parameters are given in Table 1 for the optical multi-span fibre link under consideration, which comprises multiple standard single-mode fibre spans of $75$~km, amplified at the end of each span by an EDFA with noise figure of $5$~dB. The encoded bits are mapped according to four modulation formats: PS-16QAM, constant-modulus 7b4D-2A8PSK, and 4D-OS128. 
Each of the 11 WDM channels carries independent data, where all of them are assumed to have the same transmitted power. An ideal receiver is used for detection and chromatic dispersion is digitally compensated.

\begin{figure}[!tb]
\includegraphics[width=0.48\textwidth]{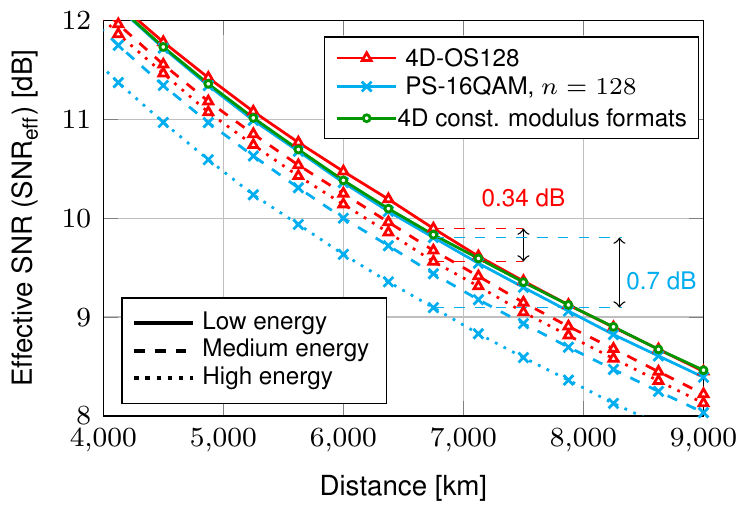}
\caption{$\text{SNR}_{\text{eff}}$  vs. transmission  distance at $P_{\text{ch}}=-1$~dBm.}
\label{fig:4D-OS128_symbols_PvsSNR}
\vspace{-0.7em}
\end{figure}

\begin{figure*}[!b]
	\vspace{-1em}
	\subfigure[Effective SNR vs. $P_{\text{ch}}$ at 7875~km.]{\includegraphics[width=0.32\textwidth]{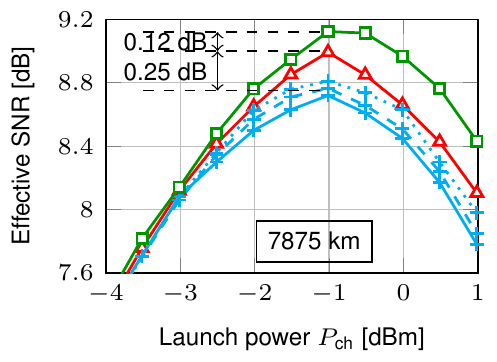}}
	\hspace{-11.5em}
	\subfigure[AIR vs. $P_{\text{ch}}$ at 7875~km.]{
	\includegraphics[width=0.87\textwidth]{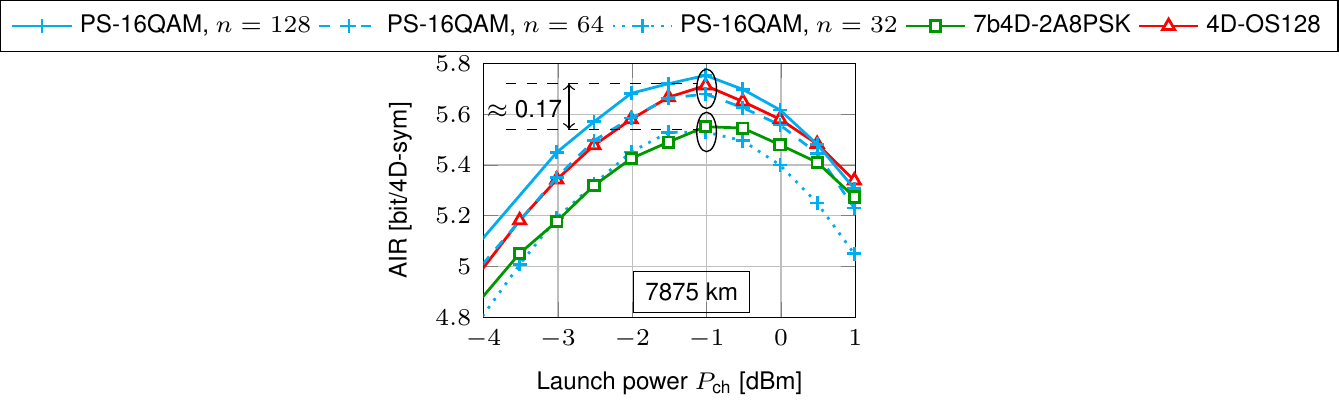}
}
	\hspace{-14.5em}
	\subfigure[AIR vs.  distance at optimal $P_{\text{ch}}$.]{
	\includegraphics[width=0.33\textwidth]{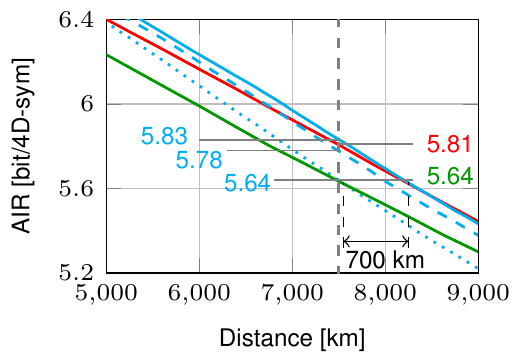}}
	\caption{Simulation results of multi-span optical fiber transmission with 11 WDM channels for three modulation formats with SE of 7~bit/4D-sym: 7b2D-4D2A8PSK, 4D-OS128 and PS-16QAM with three different blocklengths. 
	}
	\label{fig:7bit_modulation}
	\vspace{-1em}
\end{figure*}

In Fig.~\ref{fig:4D-OS128_symbols_PvsSNR}, the effective SNR ($\text{SNR}_{\text{eff}}$) in dB  versus transmission distance for three modulation formats is evaluated. The $\text{SNR}_{\text{eff}}$ is estimated from the transmitted data $X$ and received symbols $Y$ as $E_s/\sigma^2$ where $\sigma^2=\text{var}(Y-X)$ denotes the noise variance. Fig.~\ref{fig:4D-OS128_symbols_PvsSNR} shows the average SNRs calculated for the three energy levels: low, medium, and high (see Fig.~\ref{fig:4DEnergy_hist}). We observe that the $\text{SNR}_{\text{eff}}$ varies significantly with 4D symbols' energy for 4D-OS128 and PS-16QAM.  At a distance of 6750~km, the SNR difference between low and high energy symbols of PS-16QAM is $0.7$~dB, while the SNR difference for 4D-OS128 is only $0.34$~dB. The reason for this effect is that the transmitted symbols of 4D-OS128 are more close to a constant-modulus sequences than PS-16QAM's.
This observation permits us to assert that the  proposed  4D-OS128 is  more  robust than  PS solutions to fiber nonlinear impairments, and thus, gives an intuition about optimal design and implementation of future MD formats.

In Fig. \ref{fig:7bit_modulation}, we show the SNR and AIR of 7b4D-2A8PSK, 4D-OS128 and PS-16QAM with DM blocklengths $n = 32,64, 128$. We observe that PS with $n=128$ gives slightly higher AIR, but the resulting increased rate loss diminishes the efficiency of DM as the blocklength decreases.
With  $n = 32$ and $n = 64$, PS-16QAM has even worse  performance than 4D-OS128. These losses of PS-16QAM are shown in Fig.  \ref{fig:7bit_modulation} (a) and (b), 
and  are particularly  visible in the highly nonlinear regime.  {For the considered optical fiber transmission setup, 4D-OS128 can achieve approximately 9.3\% (700~km) of reach increase with respect to 7b4D-2A8PSK and PS-16QAM with blocklength $n=32$ at the same transmission rate.}
In Fig. \ref{fig:7bit_modulation} (a), we also observe that PS with short blocklengths can also slightly increase the nonlinear tolerance, and thus, the $\text{SNR}_{\text{eff}}$. The phenomena has been reported in \cite{Amari2019_IntroducingESSoptics} and very recently in \cite{FehenbergerJLT2020}.

%
%
%

\vspace{-0.5em}
\section{Conclusions}
We have studied the performance of various signal shaping schemes in the presence of fibre nonlinearities. All formats have with the same spectral efficiency (7~bit/4D-sym), however, they differ greatly in the distribution of their symbol energies. 4D symbol energy considerations showed that constant-modulus constellations reduce the NLI and that probabilistic shaping exhibits large SNR variations across symbols with different energies. The newly introduced 4D-OS128 format was shown to be able to trade-off linear and nonlinear tolerance giving SNR improvements with respect to PS-16QAM. This is achieved by introducing less 4D symbol energy variations in the transmit sequences that effectively mitigates fiber nonlinearities.

\begin{spacing}{0.8}
{\footnotesize
\linespread{0.8} \textbf{Acknowledgements}: 
The work of B. Chen is supported  by the National Natural Science Foundation of China (NSFC) under Grant 61701155.   C. Okonkwo is supported in part by the Dutch NWO Gravitation Program: Research centre for Integrated Nanophotonics (Grant Number 024.002.033). The work of A. Alvarado is supported by the Netherlands Organisation for Scientific Research (NWO) via the VIDI Grant ICONIC (project number 15685). 
}
\end{spacing}


\cleardoublepage
\printbibliography

\end{document}